\documentstyle[11pt,aaspp4]{article} 

\newcommand{\hmpc}{{h^{-1}{\rm Mpc}\;}}
\newcommand{\kps}{{\rm{km\;s^{-1}\;}}}
\newcommand{\lya}{{Ly$\alpha$}}
\newcommand{\Onot}{{\Omega_{m}}}
\newcommand{\OLam}{{\Omega_{\Lambda}}}
\newcommand{\Ob}{{\Omega_{b}}}
\newcommand{\Onu}{{\Omega_{\nu}}}

\begin{document}

\title{Constraints on Cosmological Parameters from the Ly$\alpha$
Forest Power Spectrum and COBE-DMR}
\author{John Phillips$^{1}$, David H. Weinberg$^{2}$, \linebreak Rupert A. C.
Croft$^{2,3}$, Lars Hernquist$^{3,4}$, Neal Katz$^{5}$, and Max
Pettini$^{6}$}

\footnotetext[1]{Department of Physics, The Ohio State University,
Columbus, OH 43210; phillips@pacific.mps.ohio-state.edu}

\footnotetext[2]{Department of Astronomy, The Ohio State University,
Columbus, OH 43210; dhw@astronomy.ohio-state.edu}

\footnotetext[3]{Harvard-Smithsonian Center for Astrophysics, Cambridge,
MA 02138; rcroft,lars@cfa.harvard.edu}

\footnotetext[4]{Lick Observatory, University of California, Santa Cruz,
CA 95064}

\footnotetext[5]{Department of Physics and Astronomy, University of
Massachusetts, Amherst, MA, 01003; nsk@kaka.phast.umass.edu}

\footnotetext[6]{Institute of Astronomy, Madingley Road, Cambridge,
CB3 0HA, UK; pettini@ast.cam.ac.uk}

\begin{abstract}

We combine COBE-DMR measurements of cosmic microwave background (CMB)
anisotropy with a recent measurement of the mass power spectrum at redshift
$z=2.5$ from \lya\ forest data to derive constraints on cosmological 
parameters and test the inflationary cold dark matter (CDM) scenario of 
structure formation.
By treating the inflationary spectral index $n$ as a free parameter,
we are able to find successful fits to the COBE and \lya\ forest
constraints in $\Onot=1$ models with and without massive neutrinos and in
low-$\Onot$ models with and without a cosmological constant.  Within each
class of model, the combination of COBE and the \lya\ forest
$P(k)$ constrains a parameter combination of the form 
$\Onot h^\alpha n^\beta \Omega_b^\gamma$, 
with different indices for each case.  
This new constraint breaks some of the degeneracies in cosmological
parameter determinations from other measurements of large scale
structure and CMB anisotropy.  The \lya\ forest $P(k)$ provides
the first measurement of the slope of the linear mass power spectrum
on $\sim\;$Mpc scales, $\nu=-2.25 \pm 0.18$, and it confirms a basic
prediction of the inflationary CDM scenario: an approximately
scale-invariant spectrum of primeval fluctuations ($n \approx 1$)
modulated by a transfer function that bends $P(k)$ towards $k^{n-4}$ on 
small scales.  Considering additional observational data, we find that
COBE-normalized, $\Onot=1$ models
that match the \lya\ forest $P(k)$ do not match the observed masses of
rich galaxy clusters and that low-$\Onot$ models with a cosmological
constant provide the best overall fit to the available data, even without
the direct evidence for cosmic acceleration from Type Ia supernovae. 
With our fiducial parameter choices, the flat, low-$\Omega_m$ models
that match COBE and the \lya\ forest $P(k)$ also match recent measurements
of small scale CMB anisotropy.
Modest improvements in the \lya\ forest $P(k)$ measurement 
could greatly restrict the allowable region of parameter space for CDM models, 
constrain the contribution of tensor fluctuations to CMB anisotropy,
and achieve a more stringent test of the current consensus model
of structure formation.

\end{abstract}

\section{Introduction}
\bigskip

Cosmological models based on cold dark matter (CDM) and simple versions
of inflation have had considerable success in accounting
for the origin of cosmic structure.  In this class of models, 
the primordial density fluctuations are Gaussian distributed,
and the shape of their power spectrum is determined by a small
number of physical parameters that describe the inflationary fluctuations
themselves and the material contents of the universe.
For specified cosmological parameters, the 
measurement of cosmic microwave background (CMB)
anisotropies by the COBE-DMR experiment 
(\cite{smoot92}; \cite{bennett96}) fixes the amplitude of the
matter power spectrum on large scales with an uncertainty of $\sim 20\%$
(e.g., \cite{bunn97}).
In this paper, we combine the COBE normalization with a
recent measurement of the matter power spectrum by Croft et al.\ (1999b,
hereafter CWPHK) to test the inflation+CDM scenario and constrain
its physical parameters.
A modified version of the method developed here is applied to a more
recent power spectrum measurement by Croft et al.\ (2001).

CWPHK infer the mass power spectrum $P(k)$ from measurements of \lya\ forest
absorption in the light of background quasars, at a mean absorption
redshift $z \approx 2.5$.  
The method, introduced by Croft et al.\ (1998), is based on the physical
picture of the \lya\ forest that has emerged in recent years from
3-dimensional, hydrodynamic cosmological simulations and related
analytic models (e.g., \cite{cen94}; \cite{zhang95}; \cite{hernquist96};
\cite{bi97}; \cite{hui97}).  By focusing on the absorption from diffuse
intergalactic gas in mildly non-linear structures, this method 
sidesteps the complicated theoretical problem of biased galaxy formation;
it directly estimates the linear theory mass power spectrum 
(over a limited range of scales)
under the assumption of Gaussian initial conditions.
Because the observational units are km s$^{-1}$, the CWPHK measurement 
probes somewhat different comoving scales for different cosmological
parameters: $\lambda\equiv 2\pi/k =2 - 12 \hmpc$ for 
$\Onot$=1, $\lambda=3 - 16 \hmpc$ 
for $\Onot$=0.55 and $\OLam$=0, and $\lambda=4 - 22 \hmpc$ for $\Onot$=0.4 
and $\OLam$=0.6 ($h \equiv H_0/100\;\kps\;{\rm Mpc}^{-1}$).
CWPHK determine the logarithmic slope of $P(k)$ on these
scales with an uncertainty $\sim 0.2$ and the amplitude with an
uncertainty $\sim 35\%$.  The extensive tests on simulations
in Croft et al.\ (1998) and CWPHK suggest that the statistical 
uncertainties quoted here dominate over systematic errors in the
method itself, though the measurement does depend on the assumption
of Gaussian primordial fluctuations and on the broad physical
picture of the \lya\ forest described in the references above.
For brevity, we will usually refer to the CWPHK determination 
of the mass power spectrum as ``the \lya\ $P(k)$.'' 

In the next Section, we discuss our choice of the parameter space
for inflationary CDM models.
The core of the paper is
Section 3, where we combine the COBE normalization with the \lya\ $P(k)$
to identify acceptable regions of the CDM parameter space.
We focus on four representative models: a low density ($\Onot < 1$)
open model, a low density flat model with a cosmological constant,
and Einstein-de Sitter ($\Onot=1$) models with pure CDM and with
a mixture of CDM and hot dark matter.
Because different parameters have nearly degenerate influences on
the predicted \lya\ $P(k)$, we are able to summarize our results
in terms of simple equations that constrain combinations of these parameters.
In Section 4, we consider other observational constraints that can
break these degeneracies, such as the cluster mass function, the
peculiar velocity power spectrum, the shape of the galaxy power spectrum,
and the CMB anisotropy power spectrum.
We review our conclusions in Section 5.

\section{Parameter Space for CDM Models}
\bigskip

In simple inflationary models, the power spectrum of density fluctuations
in the linear regime can be well approximated as a power law,
$P(k) \propto k^n$ (where $n=1$ is the scale-invariant spectrum),
multiplied by the square of a transfer function $T(k)$ that depends
on the relative energy densities of components with different equations
of state.  We will assume the standard radiation background
(microwave background photons and three species of light neutrinos)
and consider as other possible components cold dark matter, baryons,
a ``cosmological constant'' vacuum energy, and neutrinos with a 
non-zero rest mass in the few eV range.  Within this class of models,
the shape of the power spectrum is therefore determined by the 
parameters $n$, $\Omega_{\rm CDM}$, $\Omega_b$, $\Omega_\Lambda$,
$\Omega_\nu$, and $h$ (since $\rho_x = \Omega_x \rho_c \propto \Omega_x 
h^2$).
In place of $\Omega_b$ and $\Omega_{\rm CDM}$, we use the parameters
\begin{equation}
B \equiv \Omega_b h^2,
\label{eqn:Bdef}
\end{equation}
which is constrained by light-element abundances through 
big bang nucleosynthesis (\cite{walker91}), 
and 
\begin{equation}
\Omega_m \equiv \Omega_{\rm CDM} + \Omega_b + \Omega_\nu ,
\label{eqn:omdef}
\end{equation}
which fixes $\Omega_{\rm CDM}$ once $B$, $h$, and $\Omega_\nu$ are specified.
For non-zero $\Onu$, we assume one dominant family of massive neutrinos.
We do not consider arbitrary combinations of $\Omega_m$ and $\Omega_\Lambda$
but instead restrict our attention to the two theoretically simplest
possibilities, spatially flat models with $\Omega_\Lambda = 1-\Omega_m$
and open models with $\Omega_\Lambda=0$.

Once the cosmological parameters are specified, normalizing to the
results of the COBE-DMR experiment determines the amplitude of $P(k)$.
For inflation models with $n<1$, the COBE normalization can also be
affected by the presence of tensor fluctuations (gravity waves).
We consider normalizations with no tensor contribution and normalizations
with the quadrupole tensor-to-scalar ratio $T/S=7(1-n)$ predicted by simple
power law inflation models (e.g., \cite{davis92}), 
but we do not consider arbitrary tensor contributions.  
We compute the COBE-normalized, linear theory, matter power spectrum $P(k)$
using the convenient and accurate fitting formulas of Eisenstein \& Hu
(1999), with the normalization procedures of Bunn \& White (1997) for all 
flat cases and for the open case without a tensor contribution and Hu \& White
(1997) for the open case with a tensor contribution.                     

There are plausible variants of this family of inflationary CDM models
that we do not analyze in this paper, because we lack the tools to easily
calculate their predictions and because they would make our parameter
space intractably larger.  Prominent among these variants are models
with a time-varying scalar field, a.k.a. ``quintessence'' (e.g.,
\cite{peebles88}; \cite{wang98}), models in which the energy of 
the radiation background has been boosted above its standard value
by a decaying particle species, a.k.a. ``$\tau$CDM'' (e.g., \cite{bond91}),
and models in which inflation produces a power spectrum with 
broken scale invariance (e.g., \cite{kates95}).
Given the observational evidence for a negative pressure component 
from Type Ia supernovae (\cite{riess98}; \cite{perlmutter99}), the
quintessence family might be especially interesting to explore in
future work.

In sum, the free parameters of our family of cosmological models are
$\Omega_m$, $h$, $n$, $B$, $\Omega_\nu$, $\Omega_\Lambda$, and $T/S$.
We allow $\Omega_m$, $h$, $n$, $B$, and $\Omega_\nu$ to assume a 
continuous range of values.  For $\Omega_\Lambda$ and $T/S$ we consider
only two discrete options: $\Omega_\Lambda=0$ or $1-\Omega_m$,
and $T/S=0$ or $7(1-n)$.  

\section{Cosmological Parameters and the \lya\ Forest P(k)}
\bigskip

To organize our discussion and guide our analysis, we focus on 
variations about four fiducial models, each motivated by a combination of 
theoretical and observational considerations. The fiducial models 
are a flat cold dark matter model with a non-zero cosmological constant 
({$\Lambda$}CDM), 
an open cold dark matter model with no cosmological 
constant (OCDM), 
an $\Omega_m=1$ cold dark matter model with 
a significantly ``tilted'' inflationary spectrum (TCDM), 
and an $\Omega_m=1$ model with a mixture of cold and hot dark matter
(CHDM).

For all of the fiducial models we adopt $B=0.02$, based on
recent measurements
of the deuterium abundance in high-redshift Lyman limit absorbers
(\cite{burles97}, 1998).
For the TCDM and CHDM models we adopt $h=0.5$ in order to obtain
a reasonable age for the universe given the assumption that $\Onot=1$.
For the $\Lambda$CDM and OCDM models we instead adopt $h=0.65$, which is 
better in line with recent direct estimates of the Hubble constant
(e.g., \cite{mould00}).
For the $\Lambda$CDM model we take $\Onot=0.4$, but for OCDM we
adopt a rather high density, $\Onot=0.55$, in anticipation of our
results in Section 4, where we consider the cluster mass function
as an additional observational constraint.
For the CHDM model, we take $\Omega_\nu=0.2$ and assume
one dominant species of massive neutrino; for all other models
$\Omega_\nu=0.$
With $B$, $h$, $\Omega_m$, and $\Omega_\nu$ fixed,
we are left with one free parameter, the inflationary spectral index $n$,
which we choose in order to fit the amplitude of the \lya\ $P(k)$ while 
maintaining the COBE normalization.
The required value of $n$ is different for models with no tensor
contribution to CMB anisotropies than for models with tensor fluctuations;
we refer to the fiducial models with tensor fluctuations as 
$\Lambda$CDM2, OCDM2, and TCDM2.
Because a value $n>1$ is required for CHDM and the assumption
that $T/S=7(1-n)$ therefore cannot be correct in this case, we do not 
consider a CHDM model with tensor fluctuations.
Table \ref{tbl:andrew} lists the parameters of the fiducial models.
For later reference, Table~\ref{tbl:andrew} also lists each model's value of
$\sigma_8$, the rms linear theory mass fluctuation in spheres
of radius $8\hmpc$ at $z=0$.

\begin{table}
\caption{Fiducial Models} \label{tbl:andrew}
\begin{center}\scriptsize
\begin{tabular}{crrrrrrrrrrcr}
Model & $\Onot$ & $\OLam$ & $h$ & $n$ & $B$ & $\Onu$ &
T/S& $\sigma_8$ \\
\tableline
${\Lambda}$CDM  &       0.4     &       0.6     &       0.65      &
0.96    &       0.02    &       0.0     &       0 	&	0.91 \\
${\Lambda}$CDM2 &       0.4     &       0.6     &       0.65      &
0.98    &       0.02    &       0.0     &       $7(1-n)$&	0.89 \\
OCDM    &       0.55    &       0.0     &       0.65      &       0.88
&       0.02    &       0.0     &       0 	&	0.67 \\
OCDM2   &       0.55    &       0.0     &       0.65      &       0.92
&       0.02    &       0.0     &       $7(1-n)$&	0.64 \\
TCDM    &       1.0     &       0.0     &       0.50      &       0.84 
&       0.02    &       0.0     &       0	&	0.77 \\
TCDM2   &       1.0     &       0.0     &       0.50      &       0.89
&       0.02    &       0.0     &       $7(1-n)$&	0.73 \\
CHDM    &       1.0     &       0.0     &       0.50      &       1.10
&       0.02    &       0.2     &       0 	&	0.96 \\

\end{tabular}
\end{center}
\end{table}

\begin{figure*}
\centerline{
\epsfxsize=4.5truein
\epsfbox[65 165 550 730]{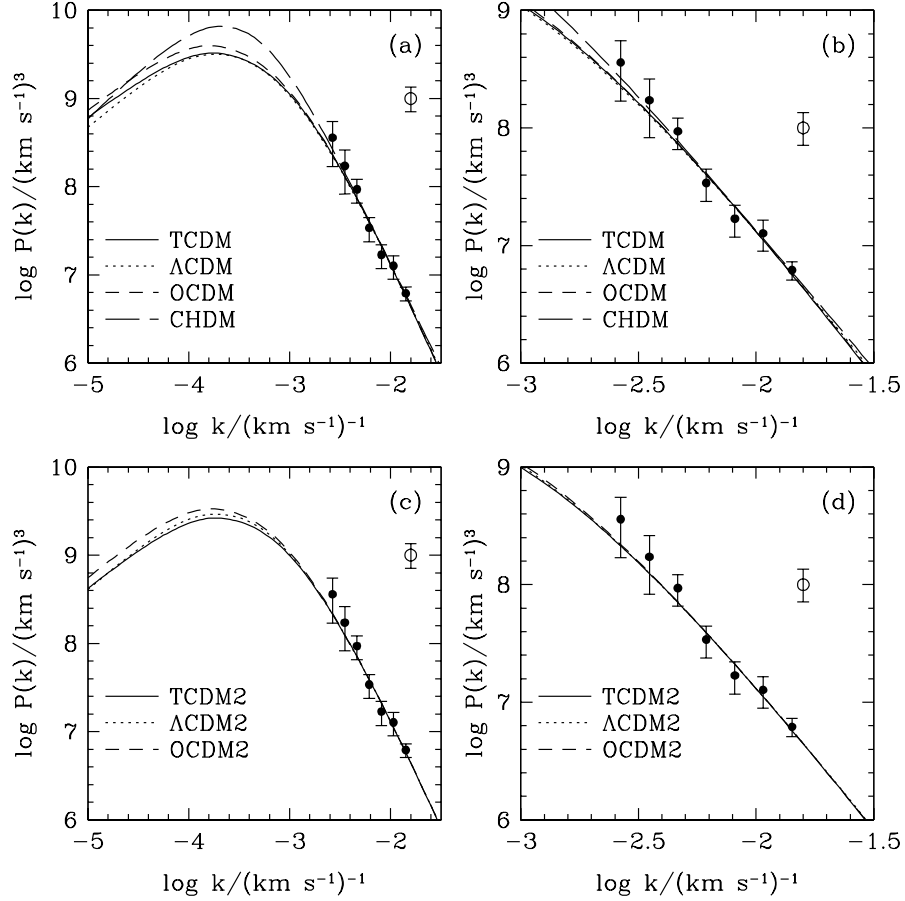}
}
\caption{
\label{fig:albert}
Power spectra of the fiducial models (smooth curves) compared to 
the \lya\ $P(k)$ determined by CWPHK (filled circles).
Upper panels show models with no tensor fluctuations and lower
panels show models with tensor fluctuations; the right hand
panels zoom in on the range of wavenumbers probed by the \lya\ $P(k)$.
The error bar on the open circle shows an overall normalization
uncertainty in the \lya\ $P(k)$; at the $1\sigma$ level all of
the points can shift up or down coherently by this amount.
Model parameters are listed in Table \ref{tbl:andrew}.
}
\end{figure*}

Figure \ref{fig:albert} compares the power spectra of our fiducial
models to the \lya\ $P(k)$, shown as the filled circles with error bars.
Note that the overall normalization of the data points is uncertain;
at the $1\sigma$ level they can shift up or down coherently by the
amount indicated by the error bar on the open circle (see CWPHK for details).
The COBE normalization itself has a $1\sigma$ uncertainty of approximately
20\% in $P(k)$, roughly half of the \lya\ $P(k)$ normalization uncertainty.
Panels (a) and (c) show the fiducial models with and without tensors,
respectively, over a wide range of wavenumber.
Panels (b) and (d) focus on the range of wavenumbers probed by the
\lya\ $P(k)$.
Our first major result is already evident from Figure \ref{fig:albert}:
all of the fiducial models reproduce the observed \lya\ $P(k)$.
Each model has a single adjustable parameter, the spectral index $n$,
so their success in reproducing both the amplitude and slope of $P(k)$
is an important confirmation of a generic prediction of the inflationary
CDM scenario, a point we will return to shortly.

Within the precision and dynamic range of the CWPHK measurement,
the \lya\ $P(k)$ can be adequately described by a power law.
CWPHK find
\begin{equation}
\Delta^{2}(k) \equiv \frac{k^{3}}{2\pi^{2}} P(k) = \Delta^{2}(k_{p}) 
\left(\frac{k}{k_{p}}\right)^{3+\nu},
\label{eqn:deltanu}
\end{equation}
with
\begin{eqnarray}
k_{p}&=&0.008\;(\kps)^{-1}, \\
\Delta^{2}(k_{p})&=&0.573^{+0.233} _{-0.166}, \\
\nu &=&-2.25\pm 0.18.
\label{eqn:values}
\end{eqnarray}
Here $\Delta^2(k)$ is the contribution to the density variance
per unit interval of $\ln k$, and $k_p$ is a ``pivot'' wavenumber
near the middle of the range probed by the data.

\begin{figure*}
\centerline{
\epsfxsize=4.5truein
\epsfbox[65 25 560 720]{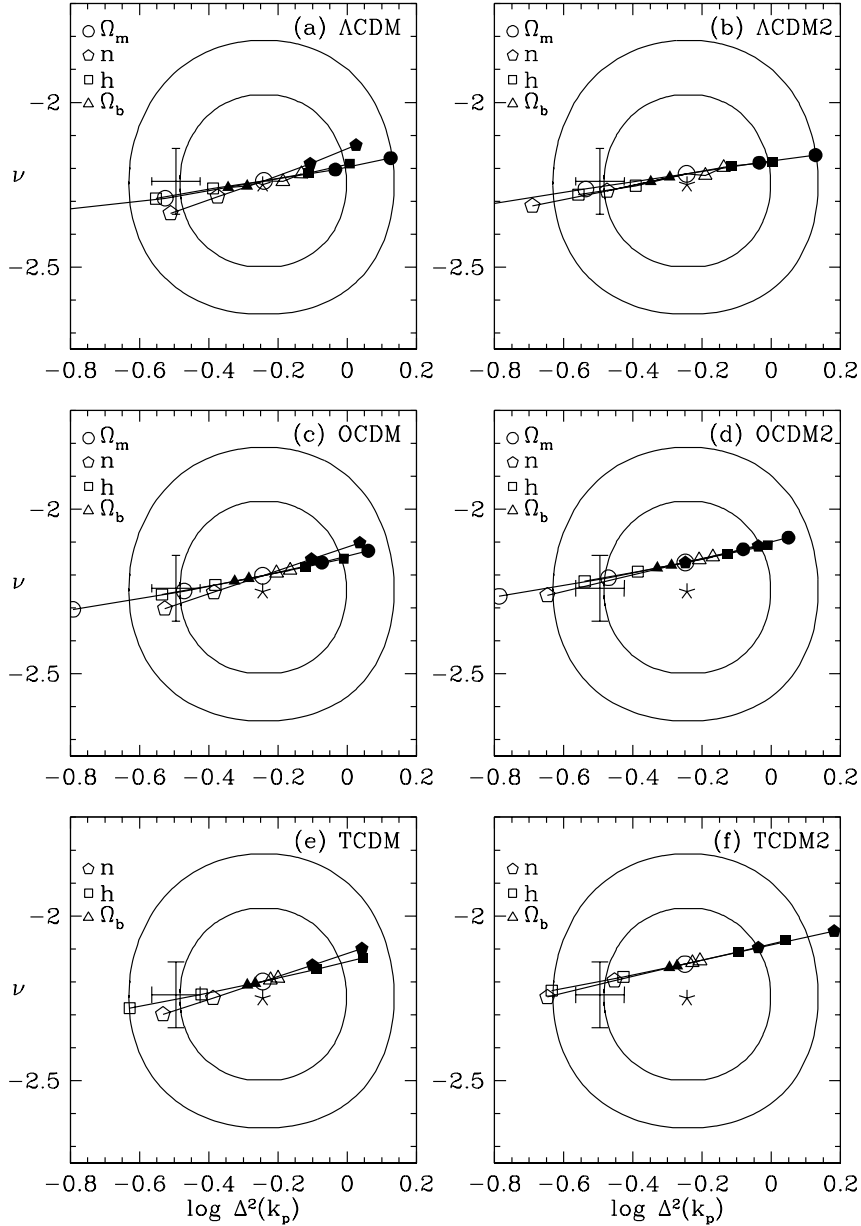}
}
\caption{
\label{fig:bertrand}
Constraints on the parameters of CDM models from COBE and the \lya\ $P(k)$
measurement.  In each panel, the central star shows CWPHK's best-fit
values of $\Delta^2(k_p)$ and $\nu$, and closed contours 
show the 68\% and 95\% confidence regions.
Each panel corresponds to a different one of the fiducial models,
with the central open circle marking the model prediction for the
parameters listed in Table~\ref{tbl:andrew}.
Other filled (open) points show the effects of increasing (decreasing)
these parameters by fixed amounts while keeping all other parameters fixed.
Circles show changes $\Delta \Onot = 0.1$, pentagons $\Delta n=0.05$,
squares $\Delta h = 0.05$, and triangles $\Delta \Ob=0.01$;
$\Omega_m$ changes are not considered for TCDM, and $n>1$ is not
considered for models with tensor fluctuations.
The error cross shows the \lya\ $P(k)$ measurement of McDonald et al.\ (2000).
}
\end{figure*}

\begin{figure*}
\centerline{
\epsfxsize=4.00truein
\epsfbox[0 405 385 785]{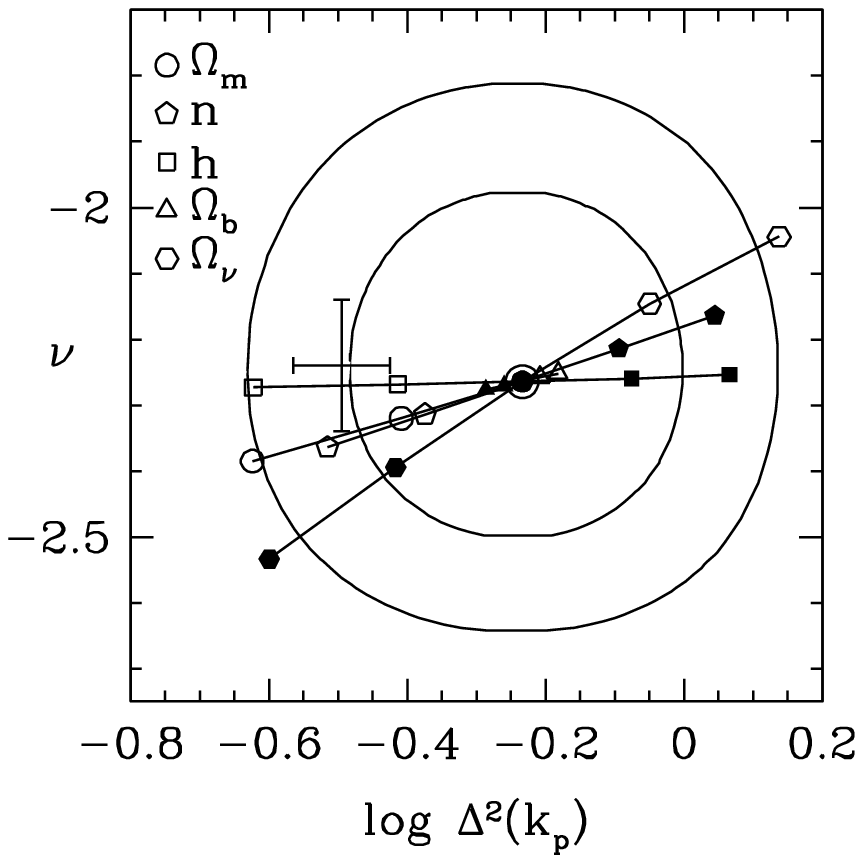}
}
\caption{
\label{fig:calvin}
Like Figure \ref{fig:bertrand}, but for the CHDM model.  Hexagons show 
changes of $\Onu$ by $\Delta \Onu=0.1$.
}
\end{figure*}

In each panel of Figures~\ref{fig:bertrand} and~\ref{fig:calvin},
the central star shows the best fit values of $\Delta^2(k_p)$ and $\nu$
quoted above, and the two large concentric circles show the $1\sigma$ (68\%)
and $2\sigma$ (95\%) confidence contours on the parameter values.
The calculation of these confidence contours is described in
detail in Section 5 of CWPHK. Briefly, the likelihood
distribution for the slope, $\nu$,
is derived by fitting the power law
form (eq.~\ref{eqn:deltanu}) to the $P(k)$ data points,
using their covariance matrix. The likelihood 
distribution for the amplitude, $\Delta^2(k_p)$,
is obtained by convolving the distributions calculated
from two separate sources of uncertainty involved in the $P(k)$
normalization.  The joint confidence contours on the two parameters 
are obtained by multiplying together the two independent 
likelihood distributions. The  $1\sigma$ and
$2\sigma$ contours correspond to changes in  $-2\log_{e}\cal{L}$
from its best fit value of $2.30$ and $6.17$, respectively,
where $\cal{L}$ is the likelihood.

The open circular point near the middle of each panel of these figures
shows the fiducial model's prediction of $\Delta^2(k_p)$ and $\nu$.
$\Lambda$CDM, OCDM, and TCDM models without tensors appear in the left 
column of Figure~\ref{fig:bertrand}, the corresponding models
with tensors appear in the right column of Figure~\ref{fig:bertrand},
and the CHDM model appears in Figure~\ref{fig:calvin}.
As expected from Figure~\ref{fig:albert}, the fiducial model
predictions lie well within the 68\% confidence contour in all cases.
The 20\% COBE normalization uncertainty adds a $\log(1.2)\approx 0.08$
error bar to the predicted value of $\log\Delta^2(k_p)$, which we have not
included in the plots.  Because this uncertainty is small (once added
in quadrature) compared to the \lya\ $P(k)$ uncertainty itself, we
have ignored it in the analysis of this paper.  With a higher precision
\lya\ forest measurement, it would be important to include the COBE
normalization uncertainty as an additional source of statistical error.

Changing any of the parameter values in any of the models shifts
the predicted $\Delta^2(k_p)$ and $\nu$, and the remaining points
in Figures~\ref{fig:bertrand} and~\ref{fig:calvin} show the
effects of such parameter changes.  Taking the $\Lambda$CDM model
of Figure~\ref{fig:bertrand}a as an example, the two filled circles
show the effect of increasing $\Onot$ by 0.1 and 0.2 
(to $\Onot=0.5$ and $\Onot=0.6$), while maintaining
the condition $\Onot+\OLam=1$ and keeping all other parameters
fixed at the fiducial values listed in Table~\ref{tbl:andrew}.
The two open circles show the effect of decreasing $\Onot$ by 0.1 and 0.2.
With $\Onot=0.2$ and other parameters unchanged (leftmost open circle),
the predicted amplitude
$\Delta^2(k_p)$ falls below the 95\% confidence lower limit of CWPHK.
In similar fashion, filled (open) pentagons show the effect of
increasing (decreasing) $n$ by 0.05, filled (open) squares show the
effect of increasing (decreasing) $h$ by 0.05, and filled (open)
triangles show the effect of increasing (decreasing) $\Omega_b$ by 0.01,
in all cases keeping the other parameters fixed at their fiducial values.
The format of the other panels of Figure~\ref{fig:bertrand} is identical, 
except that we do not show $\Onot$ changes for TCDM.
For $\Lambda$CDM2 and OCDM2, we do not allow $n>1$.
In Figure~\ref{fig:calvin}, filled (open) hexagons show the effect
of increasing (decreasing) $\Omega_\nu$ by 0.1 while keeping
$\Onot=1$.  Open circles show the effect of decreasing $\Onot$ by 0.1
while adding $\OLam$ to maintain flat space; results are virtually
indistinguishable if $\OLam$ is zero and the universe becomes (slightly)
open.  We do not consider changes that make $\Onot>1$.

Parameter changes have similar effects in all of the models, and
these effects can be easily understood by considering the physics
that determines the shape and normalization of the matter power spectrum.
The CDM transfer function has a single fundamental scale $ct_{\rm eq}$ 
determined by the size of the horizon at the time of matter-radiation 
equality; this scale
is roughly the wavelength at which the power spectrum turns over.
Increasing $h$, and hence the matter density $\rho_m\propto \Onot h^2$,
moves matter-radiation equality to higher redshift and lower $t_{\rm eq}$,
shifting the model power spectrum towards smaller scales (to the
right in Figure~\ref{fig:albert}).
This horizontal shift, combined with an upward vertical shift to
maintain the COBE normalization on large scales,
increases the amplitude of $P(k)$ on \lya\ forest scales and
translates a shallower (higher $\nu$) part of the spectrum to $k_p$.
Increasing $\Onot$ also lowers $t_{\rm eq}$ and therefore has a similar 
effect.  Open models are more sensitive than flat models 
to changes in $\Onot$ 
because the integrated Sachs-Wolfe effect makes a greater
contribution to large scale CMB anisotropies (\cite{sachs67}; \cite{hu97a}).
Increasing $\Onot$ reduces the
integrated Sachs-Wolfe effect and hence increases the matter fluctuation
amplitude implied by COBE, shifting the power
spectrum vertically upward.  The value of $\Delta^2(k_p)$ is 
sensitive to the spectral index $n$ because of the very long lever
arm between the COBE normalization scale and the scale of the 
\lya\ forest measurement.  A small decrease in $n$ produces an equally
small decrease in $\nu$ but a large decrease in $\Delta^2(k_p)$.
The fluctuation amplitude is even more sensitive to $n$ in tensor models
because, with $T/S=7(1-n)$, decreasing $n$ also increases the contribution
of gravity waves to the observed COBE anisotropies and therefore
reduces the implied amplitude of the (scalar) matter fluctuations.
Since fluctuations in the baryon component can only grow after the
baryons decouple from the photons, increasing $B$ depresses and steepens
$P(k)$ on small scales and therefore reduces $\Delta^2(k_p)$ and $\nu$.
However, for our adopted parameters the baryons always contribute a small
fraction of the overall mass density, so the influence of $\Omega_b$
changes is small.  Increasing $\Omega_\nu$ in the CHDM model has
a much greater effect in the same direction, since the suppression of
small scale power by neutrino free streaming is much greater than
the suppression by baryon-photon coupling.

Figures~\ref{fig:bertrand} and~\ref{fig:calvin} re-emphasize the
point made earlier in our discussion of Figure~\ref{fig:albert}:
the agreement between the predicted and measured slope of the
\lya\ $P(k)$ confirms a general prediction of the inflationary CDM scenario.
Although the four fiducial models correspond to quite different
versions of this scenario, all of them reproduce the measured 
value of $\nu=-2.25$ to well within its $1\sigma$ uncertainty
once the value of $n$ is chosen to match the measured $\Delta^2(k_p)$.
However, if the measured value of $\nu$ had been substantially
different, e.g.\ implying $\nu > -2$ or $\nu < -2.5$,
then none of these models could have reproduced 
the measured $\nu$ while remaining consistent with the
measured $\Delta^2(k_p)$, even allowing changes in $n$, $\Onot$, $h$,
$\Onu$, or $\Ob$. 
A different value of $\nu$ would therefore have been a challenge to the
inflationary CDM scenario itself rather than to any specific version of it.
Note also that any of the models would match the observed $\nu$ within
its $1\sigma$ uncertainty even if we had assumed a scale-invariant,
$n=1$ inflationary spectrum; it is the $\Delta^2(k_p)$ measurement
that requires the departures from $n=1$.  Because of the long
lever arm from COBE to the \lya\ $P(k)$, parameter changes that have
a modest effect on $\nu$ have a large effect on $\Delta^2(k_p)$.

Figure~\ref{fig:bertrand} also shows that changes of the different model
parameters have nearly degenerate effects on the predicted
values of $\Delta^2(k_p)$ and $\nu$. 
For example, in the $\Lambda$CDM model, 
increasing $\Onot$ by 0.1 would increase the 
predicted slope and amplitude, but decreasing $h$ by 0.05 would
almost exactly cancel this change. 
This near degeneracy allows us to summarize the constraints
imposed by COBE and the \lya\ $P(k)$ with simple formulas of the form
\begin{equation}
\Onot h^{\alpha}n^{\beta}B^{\gamma}=k\pm \epsilon,
\label{eqn:constraint1}
\end{equation}
where $k$ is the value obtained for the best-fit parameter values
in Table~\ref{tbl:andrew} and the uncertainty $\epsilon$ defines
the variation that is allowed 
before the model leaves the 68\% confidence contour.
Table~\ref{tbl:barnard} lists the values of $\alpha$, $\beta$,
$\gamma$, $k$, and $\epsilon$ for all of the fiducial models.
Although we do not show $\Onot$ changes for the TCDM models in
Figure~\ref{fig:bertrand}, we vary it below $1.0$ (adding $\OLam$
to keep the universe flat) in order to derive the $\alpha$, $\beta$,
and $\gamma$ indices, so that in all models their values reflect the importance
of a change in $h$, $n$, or $B$ relative to a change in $\Onot$.

\begin{table}
\caption{Constraint Parameters (see equation~\ref{eqn:constraint1})} 
\label{tbl:barnard}
\begin{center}\scriptsize
\begin{tabular}{crrrrrrrrrrr}
Model	&	$\alpha $	&	$\beta $	&	$\gamma
$
&	$\delta $	&	$k$	&	$\epsilon $ \\
\tableline
${\Lambda}$CDM	&	1.88	&	2.68	&	-0.26	&
--	&	0.44	&	0.12 \\
${\Lambda}$CDM2	&	1.84	&	4.48	&	-0.25	&
--	&	0.43	&	0.08 \\
OCDM	&	1.55	&	2.57	&	-0.23	&	--
&	0.50	&	0.10 \\
OCDM2	&	1.80	&	3.45	&	-0.18	&	--
&	0.38	&	0.08 \\
TCDM	&	2.33	&	4.08	&	-0.40	&	--
&	0.46	&	0.14 \\
TCDM2	&	1.82	&	4.60	&	-0.15	&	--
&	0.30	&	0.08 \\
CHDM	&	0.93	&	1.74	&	-0.13	&	-0.37
&	1.87	&	0.26 \\

\end{tabular}
\end{center}
\end{table}

Equation~(\ref{eqn:constraint1}), together  with 
Table~\ref{tbl:barnard}, is our second principal result,
defining the quantitative constraints placed on the parameters of
inflationary CDM models by the combination of COBE and the \lya\ 
forest $P(k)$.  The values of the $\alpha$, $\beta$, and $\gamma$
indices reflect the sensitivity of the predicted power spectrum
amplitude $\Delta^2(k_p)$ to the model parameters, quantifying
the impressions from Figure~\ref{fig:bertrand}.
Again taking $\Lambda$CDM as an example, we see that 
small variations in $h$ and $n$ have much greater effect 
than small variations in $\Onot$, and that the suppression of
small scale power from increases in $B$ is always a modest effect.
Models with tensors are much more sensitive to $n$ than
models without tensors because of the influence of gravity waves
on the $P(k)$ normalization, as discussed above.
Although the index
values are derived in all cases by considering small variations
about the corresponding fiducial model, the constraint 
formula~(\ref{eqn:constraint1}) remains accurate even for
fairly large changes in the cosmological parameters.
For example, plugging the TCDM values of $\Onot$, $h$, $n$, $B$
into equation~(\ref{eqn:constraint1}) with the $\Lambda$CDM
values of $\alpha$, $\beta$, and $\gamma$ yields $k=0.47$,
compared to the value $k=0.44$ listed for $\Lambda$CDM
in Table~\ref{tbl:barnard}.

Figure~\ref{fig:calvin} shows that the effects of parameter changes
are less degenerate in the CHDM model.
This difference in behavior is not surprising, since neutrino
free streaming changes $P(k)$ by depressing it at
small scales rather than simply shifting or tilting it.
The slope $\nu$ is therefore much more sensitive to changes in $\Onu$
than to changes in other parameters.  Conversely, the influence of
$h$ on $\nu$ through shifting $t_{\rm eq}$ is nearly cancelled by
the effect of $h$ on the implied neutrino mass and free streaming length.
We still analyze this case as above, adding a factor of $\Onu^{\delta}$, 
to obtain
\begin{equation}
\Onot h^{\alpha}n^{\beta}B^{\gamma}\Onu^{\delta}=k\pm\epsilon,
\label{eqn:constraint2}
\end{equation}
with parameters also listed in Table~\ref{tbl:andrew}.
However, this equation cannot describe the results of Figure~\ref{fig:calvin}
as accurately as equation~(\ref{eqn:constraint1}) describes 
the results of Figure~\ref{fig:bertrand}.

Recently McDonald et al.\ (2000)
measured the \lya\ forest flux
power spectrum in a sample of eight Keck HIRES spectra and used
it to infer the amplitude and shape of the mass power spectrum.
Their mean absorption redshift is $z \approx 3$ rather than $z=2.5$,
and their data best constrain the $P(k)$ amplitude at
$k=0.04\;(\kps)^{-1}$ rather than $0.008\;(\kps)^{-1}$.
However, assuming gravitational instability and a CDM power spectrum
shape, they extrapolate from their result to derive values of $\nu$
and $\Delta^2$ that can be directly compared to CWPHK's measurement
at $z=2.5$, $k_p=0.008\;(\kps)^{-1}$, obtaining $\nu=-2.24 \pm 0.10$ and
$\Delta^2(k_p)=0.32 \pm 0.07$.  Despite the entirely independent
data sets and very different modelling procedures, the CWPHK and
McDonald et al.\ (2000) measurements agree almost perfectly in slope
and are consistent in amplitude at the $\sim 1\sigma$ level.
We plot the McDonald et al.\ (2000) measurement 
as error crosses in Figures~\ref{fig:bertrand} and~\ref{fig:calvin}.
McDonald et al.\ (2000) note that the small error bar on $\Delta^2$ should
be considered preliminary, since they have not fully investigated the
sensitivity of their power spectrum normalization procedure to their
modelling assumptions. 

Clearly none of our qualitative conclusions about inflationary CDM
models would change if we were to adopt the McDonald et al.\ (2000)
$P(k)$ determination instead of the CWPHK determination.
Conveniently, the McDonald et al.\ (2000) point lies almost exactly
on our $-1\sigma$ error contour, so to a good approximation one
can obtain the parameter 
constraints~(\ref{eqn:constraint1}) and~(\ref{eqn:constraint2})
implied by the McDonald et al.\ (2000) measurement
by simply replacing the values of $k$ in Table~\ref{tbl:barnard}
by $k-\epsilon$.

\section{Combining with other constraints}

We have shown that the combination of COBE and the \lya\ $P(k)$ 
yields constraints on degenerate combinations of cosmological parameters. 
To break these degeneracies, we now consider observational 
constraints from other studies of large scale structure and CMB anisotropies.
Analyses of cosmological parameter constraints from multiple observations
have been carried out by numerous groups
(recent examples include \cite{bahcall99}; \cite{bridle99};
\cite{steigman99}; \cite{novosyadlyj00}; \cite{wang01}).
Our new contribution is to include the \lya\ $P(k)$ as one of
the observational constraints
(also considered by \cite{novosyadlyj00} and \cite{wang01}).
We focus our attention on several other constraints that can be cast into a 
form that complements our results from Section 3: the mass function
of galaxy clusters, 
the mass power spectrum inferred from galaxy peculiar velocities,
the shape parameter of the galaxy power spectrum, 
and a constraint on $n$ from CMB anisotropy data.
Our discussion in this Section will be more qualitative than our
discussion in Section 3, in part because the uncertainties in these
constraints are largely systematic, so that a straightforward statistical
combination could be misleading.

\begin{figure*}
\centerline{
\epsfxsize=3.7truein
\epsfbox[65 35 560 720] {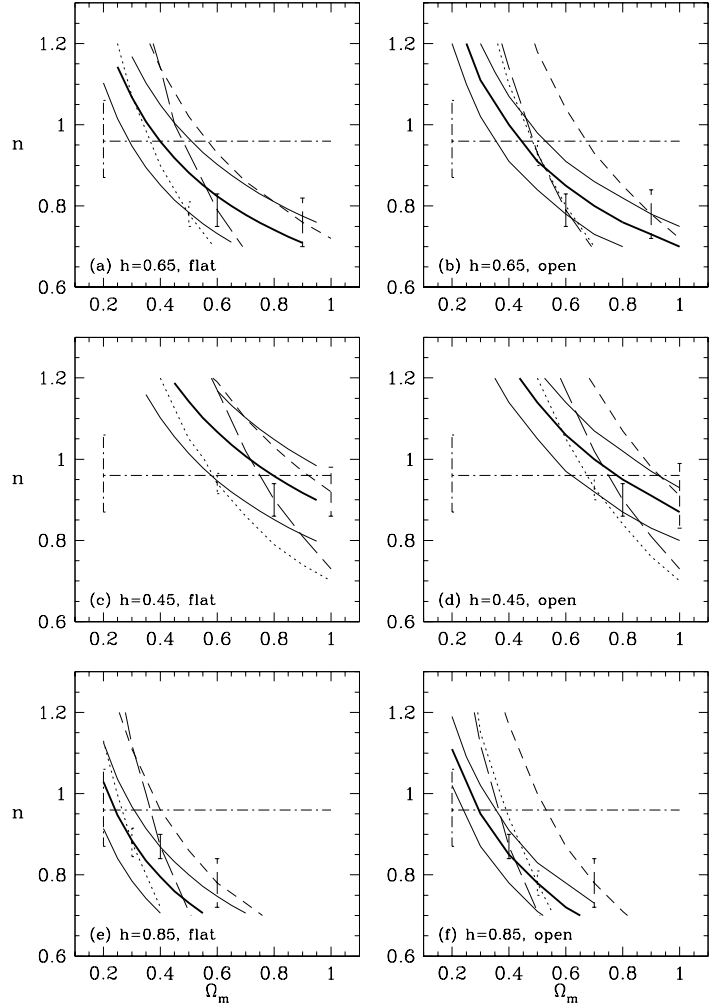}
}
\caption{
\label{fig:delmer}
Constraints in the $\Onot-n$ plane from a variety of cosmological tests,
for models with no tensor fluctuations.
Upper, middle, and lower panels show models with $h=0.65$, $h=0.45$,
and $h=0.85$, respectively, and in all cases we keep $B=0.02$.
Flat models appear in the left hand column, open models with $\OLam=0$
in the right hand column.
In each panel, the heavy solid line shows the $\Onot-n$ locus determined
by the combination of COBE and the \lya\ $P(k)$, and the light solid
lines show the $\pm 1\sigma$ range of this locus.
Dotted lines show the constraint~(\ref{eqn:sigma8}) from the cluster 
mass function, short-dashed lines the constraint~(\ref{eqn:velPk}) 
from the peculiar velocity power spectrum, long-dashed lines the
constraint~(\ref{eqn:gammaconst}) from the shape of the galaxy power spectrum,
and horizontal dot-dashed lines the constraint on $n$ from CMB
anisotropy measurements.  Error bars show representative $1\sigma$
statistical uncertainties in these constraints.
A model is consistent with multiple constraints if it lies in the
region of the $\Onot-n$ plane where these constraints overlap
within their uncertainties.
}
\end{figure*}

\begin{figure*}
\centerline{
\epsfxsize=3.7truein
\epsfbox[65 35 560 720] {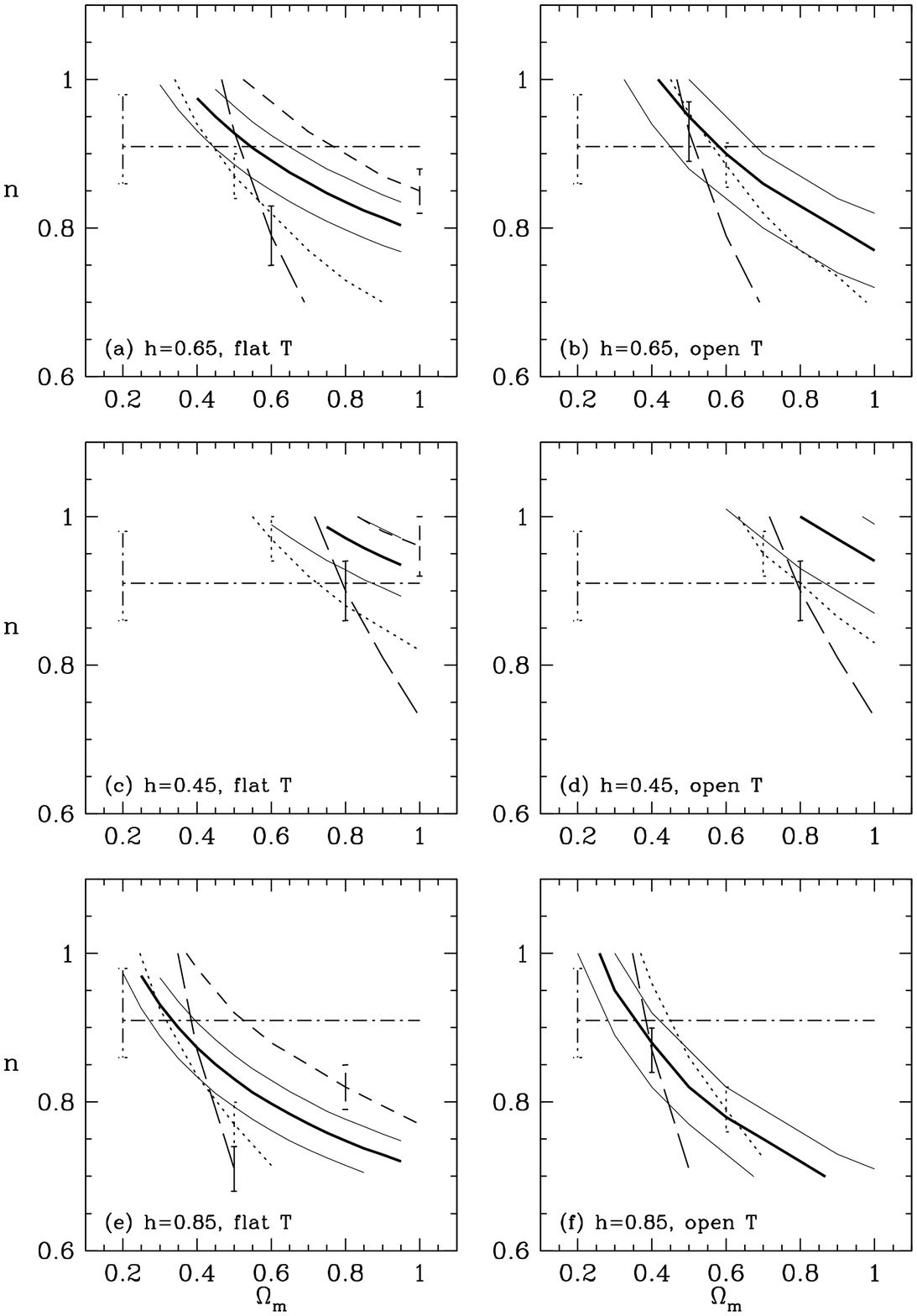}
}
\caption{
\label{fig:elbret}
Like Figure~\ref{fig:delmer}, but for models with tensor fluctuations.
Note that there is no version of the velocity power spectrum constraint
for OCDM2.  The CMB constraint on $n$ does not impose the condition
$T/S=7(1-n)$, but all of the other constraints do assume this condition
and therefore cannot be applied for $n > 1$.
} 
\end{figure*}

In each panel of Figures~\ref{fig:delmer} and~\ref{fig:elbret},
the heavy solid line shows the locus of $(\Onot,n)$ values that
yield a simultaneous match to COBE and the CWPHK measurement of
the \lya\ $P(k)$.  These lines are very close to those implied
by equation~(\ref{eqn:constraint1}) and Table~\ref{tbl:barnard},
but since those results are, strictly speaking, expansions about
our fiducial model parameters, we compute the best-fit value of $n$
exactly for each $\Onot$ rather than using equation~(\ref{eqn:constraint1}).
The $\pm 1\sigma$ constraints are shown as the lighter solid lines;
these are close to the curves implied by equation~(\ref{eqn:constraint1})
and Table~\ref{tbl:barnard} with $k$ replaced by $k \pm \epsilon$.
Because the \lya\ $P(k)$ constraint is not very sensitive to $B$,
we keep $B$ fixed at our fiducial value of 0.02 in all cases.
We show results for $h=0.65$, $h=0.45$, and $h=0.85$ in the upper,
middle, and lower panels of each figure, with flat and open models
in the left and right hand columns, respectively. 
Figure~\ref{fig:delmer} shows models without tensors and
Figure~\ref{fig:elbret} models with tensors.
For models with tensors, we restrict the parameter space to $n\leq 1$,
since our assumption that $T/S=7(1-n)$ only makes sense in this regime.
The TCDM models can be considered as the limit of either the flat
or open models at $\Onot=1$.  Note that the McDonald et al. (2000)
estimate of the \lya\ $P(k)$ corresponds very closely to our $-1\sigma$
constraint, so to adopt McDonald et al.\ (2000) instead of CWPHK
one simply follows the lower solid line instead of the middle solid line
as the constraint.

For Gaussian initial conditions, the space density of clusters as a
function of virial mass constrains a combination of $\Onot$ and the
mass fluctuation amplitude, since clusters of a given mass can be
formed by the collapse of large volumes in a low density universe
or smaller volumes in a higher density universe.
This constraint can be summarized quite accurately in a formula relating 
$\Onot$ to the rms mass fluctuation $\sigma_8$ 
(\cite{white93}).  We use the specific version of this formula
obtained by Eke, Cole, \& Frenk (1996, hereafter ECF) using N-body
simulations and the Press-Schechter (1974) approximation:
\begin{equation}
\begin{array}{ll}
\sigma_{8}=(0.52\pm 0.04)\Onot^{-0.46+0.10\Onot} &
\quad \OLam = 0 \\
\sigma_{8}=(0.52\pm 0.04)\Onot^{-0.52+0.13\Onot} &
\quad \OLam = 1-\Onot ~.
\end{array} 
\label{eqn:sigma8}
\end{equation}
For each value of $\Onot$, we find the value of $\sigma_8$ required
by the cluster mass function from equation~(\ref{eqn:sigma8}).
Given $h$ and $B=0.02$, we then find the value of $n$ required to
produce this value of $\sigma_8$ by numerically integrating the
CDM power spectrum.
This constraint in the $\Onot-n$ plane is shown by the dotted
line in each panel of Figures~\ref{fig:delmer} and~\ref{fig:elbret},
with an error bar that indicates the 8\% uncertainty quoted in 
equation~(\ref{eqn:sigma8}) from ECF.

For a given value of $\Onot$, the matter power spectrum can also be
estimated from the statistics of galaxy peculiar motions.
Freudling et al.\  (1998) apply a maximum likelihood 
technique to the SFI peculiar velocity catalog to constrain
COBE-normalized, inflationary CDM models for the matter power spectrum,
obtaining the constraint
\begin{equation}
\Onot h_{60}^{\mu}n^{\nu}=k\pm \epsilon ~,
\label{eqn:velPk}
\end{equation}
where $\mu$, $\nu$, $k$ and $\epsilon$ are dependent on the cosmology
and $h_{60}\equiv h/0.6$.
In a flat, $\OLam = 1-\Onot$ model with no tensor component, ($\mu$, 
$\nu$, $k$, $\epsilon$)=(1.3, 2.0, 0.58, 0.08), while if a tensor 
component is allowed they become (1.3, 3.9, 0.58, 0.08). For an open, 
$\OLam =0$ model without a tensor component they are (0.9, 1.4, 0.68, 
0.07).  Freudling et al.\  (1998) do not consider open, $\OLam =0$ cases 
with a tensor component.  For specified $h$, equation~(\ref{eqn:velPk})
yields a constraint in the $\Onot-n$ plane, shown by the short-dashed
line in the panels of Figures~\ref{fig:delmer} and~\ref{fig:elbret}.  
The associated $1\sigma$ error bars are based on the statistical uncertainties
$\epsilon$ quoted by Freudling et al.\ (1998) and listed above.
For brevity, we will refer
to these curves as the velocity power spectrum constraint,
though they represent the constraints on the density
power spectrum implied by peculiar velocites.

We do not want to use the amplitude of the galaxy power spectrum
as one of our constraints because it can be strongly affected
by biased galaxy formation.  However, a variety of analytic and
numerical arguments (e.g., \cite{coles93}; \cite{fry93}; 
\cite{mann98}; \cite{scherrer98}; \cite{narayanan00})
suggest that biased galaxy formation should not alter the {\it shape}
of the galaxy power spectrum on scales in the linear regime,
and on these scales the shape is directly related to the parameters
of the inflationary CDM cosmology.
We adopt the specific constraint found by Peacock \& Dodds (1994)
from their combined analysis of a number of galaxy clustering data sets:
\begin{equation}
\Gamma_{\rm eff} \equiv 
\Onot h \exp {\left[{-\Ob}\left(1+{\frac{\sqrt{2h}}{\Onot}}\right)\right]} 
- 0.32\left(\frac{1}{n} - 1\right)=0.255 \pm 0.017.
\label{eqn:gammaconst}
\end{equation}
For $n=1$, $\Gamma_{\rm eff}=\Gamma$, where $\Gamma$ is the shape 
parameter in the conventional parameterization of the inflationary
CDM power spectrum (\cite{bardeen86}; the influence 
of $\Ob$ is discussed by \cite{sugiyama95}).
While the effects of $\Gamma$ and $n$ on the power spectrum shape are 
different, equation~(\ref{eqn:gammaconst}) combines them in a way
that approximates their nearly degenerate influence over the range
of scales currently probed by large scale clustering measurements.
For specified $h$ and $B$, equation~(\ref{eqn:gammaconst})
becomes a constraint in the $\Onot-n$ plane.
We plot this constraint as the long-dashed line and associated error bar
in the panels of Figures~\ref{fig:delmer} and~\ref{fig:elbret}.
We should note, however, that the Peacock \& Dodds (1994) error bar
may be overoptimistic, since independent estimates of $\Gamma_{\rm eff}$
often fall outside this range.
Eisenstein \& Zaldarriaga (2001) have recently re-examined the 
spatial power spectrum inferred from the APM survey and conclude
that the 68\% confidence interval of $\Gamma$ (for $n=1$)
is $0.19-0.37$, much larger than the range implied by
equation~(\ref{eqn:gammaconst}), and Efstathiou \& Moody (2000)
favor a lower central value ($\Gamma\approx 0.12$, with $2\sigma$ range
$0.05 \leq \Gamma \leq 0.38$).
As older estimates of the galaxy power spectrum are supplanted by
results from the 2dF and Sloan galaxy redshift surveys, the $\Gamma$
parameterization itself may become an insufficiently accurate representation
of the theoretical predictions (\cite{percival01}).

A detailed consideration of constraints from smaller scale CMB anisotropy
measurements is beyond the scope of this paper,
but we do want to draw on limits that smaller scale measurements
place on the inflationary index $n$.
For the no-tensor models, we adopt the ``weak prior'' constraint
$n=0.96^{+0.10}_{-0.09}$ of Netterfield et al.\ (2001), based on data
from the BOOMERANG experiment, which we represent by the horizontal
dot-dash line and $1\sigma$ error bar in Figure~\ref{fig:delmer}.
Since Netterfield et al. (2001) do not consider models with tensor fluctuations,
we take the corresponding constraint for the tensor models in 
Figure~\ref{fig:elbret} from Wang et al. (2001).  Their model space is
less restrictive than ours because they do not impose the power-law
inflation relation $T/S=7(1-n)$, and using CMB data alone they find
only a very weak constraint on $n$.  We therefore adopt their constraint
from the combination of CMB and large scale structure data,
$n=0.91^{+0.07}_{-0.05}$, where we have reduced the 95\% confidence range
quoted in their table~2 by a factor of two to get a representative
$1\sigma$ uncertainty.

In Figures~\ref{fig:delmer} and~\ref{fig:elbret}, 
the cluster mass function, velocity power spectrum, and shape parameter
constraints tend to be roughly parallel to each other, with
the shape parameter following a somewhat different track when
tensor fluctuations are important.
The shape parameter constraint is usually compatible with the
cluster mass function constraint, at least if one allows for
the possibility that the error bar in equation~(\ref{eqn:gammaconst})
is somewhat too small.
However, the velocity power spectrum always implies a higher fluctuation 
amplitude than the cluster mass function, and the two constraints
are not consistent within their stated $1\sigma$ uncertainties for any 
combination of $\Onot$, $n$, and $h$.  
A recent analysis by Silberman et al. (2001) shows that the discrepancy
is probably a result of non-linear effects on the velocity power
spectrum, and that correcting for these yields results closer to
the cluster constraint.  We therefore regard the cluster constraint
as more reliable, and we retain the velocity power spectrum curve mainly
as a reminder of other data that can be brought to bear on these questions.

The \lya\ $P(k)$ curve cuts across the other three constraints,
requiring greater change in $\Onot$ for a given change in $n$.
The CMB anisotropy limit on $n$ cuts across all of the other
constraints.  The COBE-DMR measurement is represented implicitly 
in Figures~\ref{fig:delmer} and~\ref{fig:elbret} through its role
in the \lya\ $P(k)$ constraint, the velocity power spectrum constraint,
and the CMB anisotropy constraint on $n$.
The size of the $1\sigma$ error bars in these figures,
and the probability that at least some of them are underestimated,
prevents us from drawing sweeping conclusions.
However, Figures~\ref{fig:delmer} and~\ref{fig:elbret} do have
a number of suggestive implications if we look for models that
lie within the overlapping $1\sigma$ uncertainties of the various constraints.
Since it is not possible to satisfy the cluster mass function and
velocity power spectrum constraints simultaneously within the class
of models that we consider, the implications depend strongly on which
of these constraints we take to be more reliable.
The shape parameter implications are usually intermediate, but
significantly closer to those of the cluster mass function.

The combination of the velocity power spectrum and \lya\ $P(k)$
constraints implies a high density universe, with 
$\Omega_m \ga 1$ preferred and $\Omega_m \la 0.6$ separating the
two constraints by more than their $1\sigma$ error bars.
The \lya\ $P(k)$ constraint rules out the high values of $n$ 
that could otherwise allow low $\Omega_m$ in equation~(\ref{eqn:velPk}).
For $h \geq 0.65$, intersection of the velocity power spectrum
and \lya\ $P(k)$ constraints occurs at $n \la 0.8$, incompatible
with the CMB anisotropy constraint.  However, an $\Onot \sim 1$
universe would require a low value of $h$ in any case because
of the age constraint for globular cluster stars, and this would push
the intersection to higher $n$.
As noted earlier, the velocity power spectrum constraint shown here
is probably biased towards high $\Omega_m$ by the non-linear effects
described by Silberman et al. (2001).

If we instead adopt the cluster mass function constraint,
then consistency with the \lya\ $P(k)$ and COBE
requires $\Omega_m<1$.  For $h=0.65$, the constraints intersect
at $\Onot \sim 0.4-0.5$ in flat models and $\Onot \sim 0.5-0.6$ in
open models; increasing $h$ slightly decreases the preferred $\Onot$
and vice versa.  This conclusion --- that the combination of COBE,
the \lya\ $P(k)$, and the cluster mass function implies a low density
universe --- is the most important and robust result to emerge from this
multi-constraint analysis.

At one level, our conclusions about the matter density
come as no suprise, since we have 
already argued, in Weinberg et al.\ (1999), that consistency between
the cluster mass function and the \lya\ $P(k)$ implies $\Onot$ in 
this range independent of the COBE normalization.  However, the nature
of the argument is subtly different in this case.
In Weinberg et al.\ (1999), we considered matter power spectra
of the CDM form parameterized by $\Gamma$ (with $n=1$), and by
combining the \lya\ $P(k)$ measurement with the cluster 
constraint~(\ref{eqn:sigma8}), we found $\Onot=0.34+1.3(\Gamma-0.2)$
for flat models and $\Onot=0.46+1.3(\Gamma-0.2)$ for open models,
with $1\sigma$ uncertainties of about $0.1$.  However, the \lya\ $P(k)$
alone could not rule out the solution of high $\Onot$ and high $\Gamma$,
so Weinberg et al.'s (1999) conclusion that $\Onot < 1$ rested crucially
on the empirical evidence for $\Gamma \approx 0.2$ from the shape of the galaxy
power spectrum.  Within the class of CDM models considered here,
the combination of COBE and the \lya\ $P(k)$ determines $n$, and hence
the effective value of $\Gamma$ (eq.~\ref{eqn:gammaconst}), once $\Onot$,
$h$, and $B$ are specified.  Simultaneous consistency between COBE,
the \lya\ $P(k)$, and the cluster mass function requires low $\Onot$
{\it independent} of the galaxy power spectrum shape, thereby strengthening 
the overall argument for a low density universe, and, by the by, for a matter 
power spectrum with low $\Gamma_{\rm eff}$.  The lower limit on $\Onot$
from this combination of constraints varies with the choice of other 
parameters, but it never reaches as low as $\Onot = 0.2$ unless
$h \geq 0.85$.

For all of the models shown in Figures~\ref{fig:delmer} and~\ref{fig:elbret},
the \lya\ $P(k)$ and cluster mass function constraints intersect at
values of $n$ consistent with the CMB anisotropy constraints, provided
one takes the 1-$\sigma$ error ranges into account.  A factor of two
improvement in the precision of the \lya\ $P(k)$ measurement could greatly
restrict the range of models compatible with all three constraints,
especially if the \lya\ $P(k)$ amplitude is somewhat lower,
as McDonald et al. (2000) find.

There are, of course, numerous other constraints on cosmological parameters,
and we will briefly consider three of them: the cluster baryon fraction,
the location of the first acoustic peak in the CMB power spectrum,
and the evidence for accelerating expansion from Type Ia supernovae.
(Our focus on $h=0.65$ as a fiducial case already reflects our assessment
of the most convincing direct estimates of $H_0$.)
If one assumes that baryons are not overrepresented relative to their
universal value within the virial radii of rich clusters, then the 
combination of the measured gas mass fractions with big bang nucleosynthesis
limits on $\Omega_b$ yields an upper limit on $\Onot$ (\cite{white93b}).
Applying this argument, Evrard (1997) concludes that
\begin{equation}
\Onot \Ob^{-1}h^{-3/2} \leq 23.1 \quad \Longrightarrow \quad
\Onot \leq 0.57\left({B \over 0.02}\right)\left(h \over 0.65\right)^{-1/2},
\label{eqn:xrayconst}
\end{equation}
at the 95\% confidence level.  From Figures~\ref{fig:delmer}
and~\ref{fig:elbret} we see that models matching COBE, the \lya\ $P(k)$,
and the cluster mass function are always consistent with this limit ---
easily in the case of flat models, sometimes marginally in the case
of open models.  Models that match the velocity power spectrum 
instead of the cluster mass function are usually incompatible with this limit, 
though sometimes only marginally so.

The location of the first acoustic peak in the CMB anisotropy
spectrum is a strong diagnostic for space curvature
(e.g., \cite{doroshkevich78}; \cite{wilson81}; \cite{sugiyama92}; 
\cite{kamionkowski94}; Hu et al.\ 1997), and recent anisotropy 
measurements on degree scales favor a geometry that is close
to flat (e.g., \cite{miller99}; \cite{melchiorri00};
\cite{netterfield01}; \cite{pryke01}).
Clearly our flat universe models
are compatible with these results, as are the open universe models
that match the \lya\ $P(k)$ and the velocity power spectrum
(all of which have $\Onot$ close to one).  The open models that
match \lya\ $P(k)$ and the cluster mass function are generally
ruled out by the most recent, high precision limits on space curvature.
The Type Ia supernova
measurements of the cosmic expansion history (\cite{riess98};
\cite{perlmutter99}) add a great deal of discriminatory power,
since they constrain a parameter combination that is roughly
$\Onot-\OLam$ instead of $\Onot+\OLam$; Perlmutter et al.\ (1999)
quote $\Onot-0.75\OLam \approx -0.25 \pm 0.125$.
All of the open models miss this constraint by many $\sigma$, 
and the flat models matching the \lya\ $P(k)$ and the velocity
power spectrum fail because the values of $\Onot$ are too high.
The combination of COBE, the \lya\ $P(k)$, and the cluster mass function,
on the other hand, is compatible with the supernova results for flat
models with a cosmological constant, though it favors somewhat
higher values of $\Onot$.

We have not carried out a similar multi-constraint analysis for 
the CHDM model because 
the formulas~(\ref{eqn:gammaconst}) and~(\ref{eqn:velPk})
for the shape parameter and velocity power spectrum constraints
do not apply to it and the formula~(\ref{eqn:sigma8}) for the cluster
mass function constraint may be less accurate for non-zero $\Onu$.
However, our fiducial CHDM model, with
$\Onu=0.2$, has $\sigma_8=0.96$, with $n=1.10$. 
For $\Onu=0.3$ we obtain $\sigma_8=1.15$ ($n=1.23$), for
$\Onu=0.1$ we obtain $\sigma_8=0.81$ ($n=0.96$), and for
the TCDM model, which represents the limiting case of $\Onu=0$,
we obtain $\sigma_8=0.77$ ($n=0.84$).
All of these models are likely to violate the cluster mass
function constraint, which according to equation~(\ref{eqn:sigma8})
implies $\sigma_8=0.52 \pm 0.04$
for $\Onot=1$.  We conclude that COBE-normalized CHDM
models with $\Onot=1$, $h \approx 0.5$ cannot simultaneously
match the \lya\ $P(k)$ and the cluster mass function.
The \lya\ $P(k)$ strengthens the case against this class
of CHDM models by ruling out the low values of $n$ that would otherwise
allow them to match cluster masses (\cite{ma96}).
Of course CHDM models with $\Onot<1$ can satisfy the observational
constraints for appropriate parameter choices, and the general
problem of using CMB measurements and the \lya\ $P(k)$ to measure
$\Onu$ is discussed by Croft, Hu, \& Dav\'e (1999).
However, the possible presence of a neutrino component does not
alter our conclusion that COBE, the \lya\ $P(k)$, and the cluster
mass function together require a low density universe.

\begin{figure*}
\centerline{
\epsfxsize=4.0truein
\epsfbox[115 430 460 725] {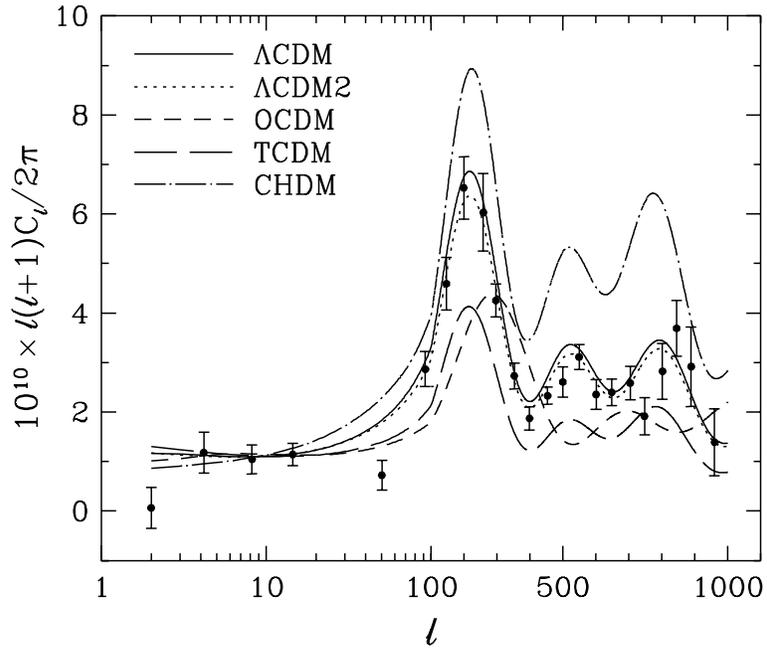}
}
\caption{
\label{fig:fagan}
CMB anisotropy power spectra for five of the fiducial models that
match COBE-DMR and the \lya\ $P(k)$.
$C_l$ is the mean-squared amplitude of spherical harmonics of order $l$.
The models shown are $\Lambda$CDM (solid), $\Lambda$CDM2 (dotted),
OCDM (long-dashed), TCDM (short-dashed), and CHDM (dot-dashed).
Parameters of the models are listed in Table~\ref{tbl:andrew}. 
Data points with $1\sigma$ errors are taken from Wang et al. (2001).
We shift from logarithmic spacing to linear spacing at $l=100$
in order to show both large and small angular scales clearly.
} 
\end{figure*}

All in all,
the CWPHK and McDonald et al.\ (2000) measurements of the \lya\ $P(k)$
provide additional support for the current ``consensus'' model of
structure formation,
$\Lambda$CDM with $\Onot \approx 0.4$ and $h \approx 0.65$.
Moderate improvements in the statistical precision of the 
constraints considered here could strengthen this support,
or they could open fissures of disagreement.
Improvements in the near future could also allow some interesting new tests,
such as discriminating between models with no tensor fluctuations
and models with the $T/S=7(1-n)$ contribution predicted by power law
inflation.

A detailed consideration of the constraints from CMB anisotropy
measurements is a major undertaking in itself, well beyond the
scope of this paper.  However, to illustrate the interplay between
our results and recent CMB experiments, we plot in Figure~\ref{fig:fagan} 
the predicted CMB power spectra of five of our fiducial models:
$\Lambda$CDM, $\Lambda$CDM2, OCDM, TCDM, and CHDM.
We computed these power spectra using CMBFAST (\cite{seljak96}; 
\cite{zaldarriaga98}), with the cosmological parameter values listed in 
Table~\ref{tbl:andrew}.  The CHDM model stands out from the rest because
matching the \lya\ $P(k)$ requires a high value of $n$, which boosts the
anisotropy on small scales.  The OCDM model also stands out,
albeit less dramatically, because the open space geometry shifts the
acoustic peaks to smaller angles.  
The TCDM model lies below the $\Lambda$CDM models because of its
larger tilt, which suppresses small scale fluctuations.
Figure~\ref{fig:fagan} shows data points taken from the joint analysis
of numerous CMB data sets by Wang et al. (2001; see their Table 1).
The two $\Lambda$CDM models fit these data points remarkably well, 
given that the choice of their parameters was not based on small
scale CMB data at all.  Because the combination of COBE and the 
\lya\ $P(k)$ implies $n$ close to one for both of these models,
their predictions are not very different, and the current CMB data 
do not distinguish between them.  However, the TCDM, OCDM, and CHDM
models are clearly ruled out, and while we have not attempted to adjust
their parameters within the constraints allowed by 
equations~(\ref{eqn:constraint1}) and~(\ref{eqn:constraint2}),
it appears unlikely that any such adjustment would allow these
models to fit the current CMB data.

\section{Conclusions}
\bigskip

The slope of the mass power spectrum inferred by CWPHK from the \lya\ forest,
$\nu=-2.25 \pm 0.18$ at $k_p=0.008\; (\kps)^{-1}$ at $z=2.5$, 
confirms one of the basic predictions of the inflationary CDM scenario:
an approximately scale-invariant spectrum of primeval 
inflationary fluctuations ($n \approx 1$) modulated by a transfer function
that bends the power spectrum towards $P(k)\propto k^{n-4}$
on small scales.  If the measured slope of the power spectrum had implied
$\nu > -2$ or $\nu < -2.5$, we would have been unable to reproduce
the \lya\ $P(k)$ with any of the models considered here, even allowing
wide variations in the cosmological parameters.

Because the amplitude of the COBE-normalized power spectrum on small
scales is very sensitive to $n$, we are able to match the 
CWPHK measurement of $\Delta^2(k_p)$ in most of the major variants
of the CDM scenario ($\Lambda$CDM, OCDM, TCDM, CHDM) by treating $n$
as a free parameter.  Within each of these variants, we obtain 
constraints on the model parameters of the form 
$\Onot h^\alpha n^\beta B^\gamma = k \pm \epsilon$ (eq.~\ref{eqn:constraint1})
or
$\Onot h^\alpha n^\beta B^\gamma \Onu^\delta = k \pm \epsilon$ 
(eq.~\ref{eqn:constraint2}), with the parameter values
listed in Table~\ref{tbl:barnard}.
These constraints, together with the confirmation of the predicted slope,
are the main results to emerge from combining the \lya\ $P(k)$ 
measurement with the COBE-DMR result.

As shown in Figures~\ref{fig:delmer} and~\ref{fig:elbret},
the parameter combination constrained by COBE and the \lya\ $P(k)$
is different from the combinations constrained by other measurements
of large scale structure and CMB anisotropy, so joint consideration
of these constraints can break some of the degeneracies among
the fundamental parameters.
If we combine the \lya\ $P(k)$ constraint with the constraint
on $\Onot$ and $\sigma_8$ inferred from the cluster mass
function (White et al.\ 1993a; ECF),
then we favor a low density universe, with $\Onot \sim 0.3-0.5$
in flat models and $\Onot \sim 0.5-0.6$ in open models.
This combination is also consistent with CMB anisotropy constraints on $n$.
The open models are inconsistent with the angular location of
the first acoustic peak in the CMB power spectrum 
(\cite{netterfield01}; \cite{pryke01}),
and they are strongly
inconsistent with Type Ia supernova results, which imply 
$\Onot-0.75\OLam \approx  -0.25 \pm 0.125$ (\cite{riess98,perlmutter99}).
The flat models are consistent
with both constraints.  On the whole, the CWPHK measurement of the
\lya\ $P(k)$ supports the consensus in favor of $\Lambda$CDM with 
$\Onot \approx 0.4$, $h \approx 0.65$.
The contribution of the \lya\ $P(k)$ to this consensus comes both
from the slope, which confirms the generic inflationary CDM prediction,
and from the amplitude, which has a different dependence on cosmological
parameters than any of the other constraints considered here.

There are bright prospects for improvements of this approach in the near
future.  McDonald et al.\ (2000) have inferred the mass power
spectrum from an independent \lya\ forest data set using a different
analysis method, obtaining a nearly identical slope and an amplitude
lower by $\sim 1\sigma$.  We have recently analyzed a much  
larger data set of high and moderate resolution spectra, using
a variant of the Croft et al.\ (1998, 1999) method, and the improved
data yield much higher statistical precision and better tests
for systematic effects.  Constraints from this new measurement of $P(k)$,
using the method developed here, are presented in \S 7 of Croft et al. (2001).
Recent measurements of CMB anisotropy have greatly improved the level
of precision on small angular scales, and results from the MAP satellite
should provide another major advance in the near future.
These measurements yield tighter cosmological parameter constraints
on their own, but they become substantially more powerful when combined
with data that constrain the shape and amplitude of the matter power 
spectrum.  It is evident from Figures~\ref{fig:delmer} and~\ref{fig:elbret}
that simply reducing the error bars on $n$ and the \lya\ $P(k)$ by a
factor of two would already produce interesting new restrictions on
the allowable range of models.  These restrictions can become very
powerful if ongoing studies of cluster masses using galaxy dynamics,
X-ray properties, the Sunyaev-Zel'dovich effect, and gravitational lensing
confirm the robustness of the cluster mass function constraint.
In the slightly longer term, the 2dF and Sloan redshift surveys should 
produce measurements of the shape of the galaxy power spectrum that
shrink the current statistical and systematic uncertainties,
so that demanding consistency between the inferred value of $\Gamma_{\rm eff}$
and other constraints becomes a useful additional test.
At the very least, these developments should lead to a powerful
test of the inflationary CDM picture and high-precision determinations
of its parameters.  If we are lucky, improved measurements
will reveal deficiencies of the simplest $\Lambda$CDM models 
that are hidden within the current uncertainties, and resolving
these discrepancies will lead us to a better understanding of 
the cosmic energy contents and the origin of primordial fluctuations
in the hot early universe.

\acknowledgments

We thank Daniel Eisenstein and Wayne Hu for helpful advice on
computing power spectra, Martin White for comments on the manuscript,
and Nikolay Gnedin for a prompt and helpful referee's report.
This work was supported by NASA Astrophysical Theory Grants NAG5-3111,
NAG5-3922, and NAG5-3820,
by NASA Long-Term Space Astrophysics Grant NAG5-3525,
and by NSF grants AST-9802568, ASC 93-18185, and AST-9803137.

\vfill\eject

\end{document}